\documentclass[12pt]{iopart}

\usepackage{graphicx}
\usepackage{cite}

\begin{document}

\title{Fano-Rashba effect in quantum dots}
\author{P A Orellana,$^1$ M Amado,$^2$ and F Dom\'{\i}nguez-Adame$^2$}

\address{$^1$ Departamento de F\'{\i}sica, Universidad Cat\'{o}lica del Norte,
Casilla 1280, Antofagasta, Chile}

\address{$^2$ GISC, Departamento de F\'{\i}sica de Materiales, Universidad
Complutense, E-28040 Madrid, Spain}

\ead{orellana@ucn.cl}

\begin{abstract}
We consider the electronic transport through a Rashba quantum dot coupled to
ferromagnetic leads. We show that the interference of localized electron
states with resonant electron states leads to the appearance of the
Fano-Rashba effect. This effect occurs due to the interference of bound
levels of spin-polarized electrons with the continuum of electronic states
with an opposite spin polarization. We investigate this Fano-Rashba effect
as a function of the applied magnetic field and Rashba spin-orbit coupling.
\end{abstract}


\submitto{\NT}

\maketitle

\section{Introduction}

Recently, there has been much interest in understanding the manner in which the
unique properties of nanostructures may be exploited in spintronic devices,
which utilize the spin degree of freedom of the electron as the basis of their
operation ~\cite{Data,Song,Folk,Mireles,Mireles2,Berciu,Zutic04}. The main
challenge in the field of spintronics is to achieve the injection, modulation
and detection of electron spin at the nanometer scale. In 1990, Datta and
Das~\cite{Data} proposed a spin transistor, based on the electron spin
precession controlled by an external electric field via spin-orbit coupling. In
this proposal, ferromagnetic contacts were used as spin-polarized source and
detector. A natural feature of these devices is the direct connection between
their conductance and their quantum-mechanical transmission properties, which
may allow their use as an all-electrical means for generating and detecting spin
polarized distributions of carriers.

Enforcing the analogy between quantum dots (QDs) and atomic systems,
Fano~\cite{Kobayashi04} and Dicke effects~\cite{Brandes} were found to be
present in several QD configurations. On the other hand, Song \emph{et
al}.~\cite{Song} described how a spin filter may be achieved in open QD systems
by exploiting Fano resonances that occur in their transmission characteristic.
In a QD in which the spin degeneracy of carrier is lifted, they showed that the
Fano effect may be used as an effective means to generate spin polarization of
transmitted carriers and that electrical detection of the resulting polarization
should be possible.

The Rashba spin-orbit interaction arises from a structure inversion asymmetry
resulting from the asymmetry of the in-plane confining potential in
semiconductor heterostructures~\cite{Rashba,Bychkov}. This effect causes a spin
splitting proportional to $k$. On the other hand, the Fano effect arises from
the interference between a localized state and the continuum. In general, the
condition for the Fano effect is the presence of two scattering channels at
least: the discrete level and continuum band. The Fano effect in electronic
transport through a single-electron transistor allows to alter the interference
between the two paths by changing the voltages on various gates. Kobayashi
\emph{et al.}~\cite{Kobayashi04} reported the first tunable Fano experiment in
which a well-defined Fano system is realized in an Aharonov-Bohm ring with a QD
embedded in one of its arms. Recently, Shelykh \emph{et al}.~\cite{Shelykh}
studied the first Fano-type resonances due to the interaction of electron states
with opposite spin orientation. They investigated the electronic transport
through Datta and Das spin-modulator devices~\cite{Data}. They show that the
interfaces make the device behave as a Fabry-Perot cavity, so that Breit-Wigner
resonances appear in the transmission coefficient. Additionally, interference of
quantum-confined electron states with free electron states leads to appearance
of asymmetric Fano line-shapes. Moreover,  S\'{a}nchez \emph{et.
al}.~\cite{sanchez,sanchez-serra} and L\'{o}pez \textit{et. al.}~\cite{lopez}
predicted the occurrence of Fano line shapes in a semiconductor quantum wire
with local spin-orbit Rashba coupling.

In this work we investigate the electronic transport through a QD with
ferro\-magnetic contacts considering the Rashba spin-orbit interaction. We show
that the interference of localized electron states with free electron states
leads to appearance of the Fano-Rashba effect. This effect appears due to the
interference of bound levels of spin-polarized electrons with the continuum of
electronic states with an opposite spin polarization. We investigate this
Fano-Rashba effect as a function of the system parameters.

\section{Model}

\begin{figure}[h]
\centerline{\includegraphics[width=7cm,angle=0,scale=1.0]{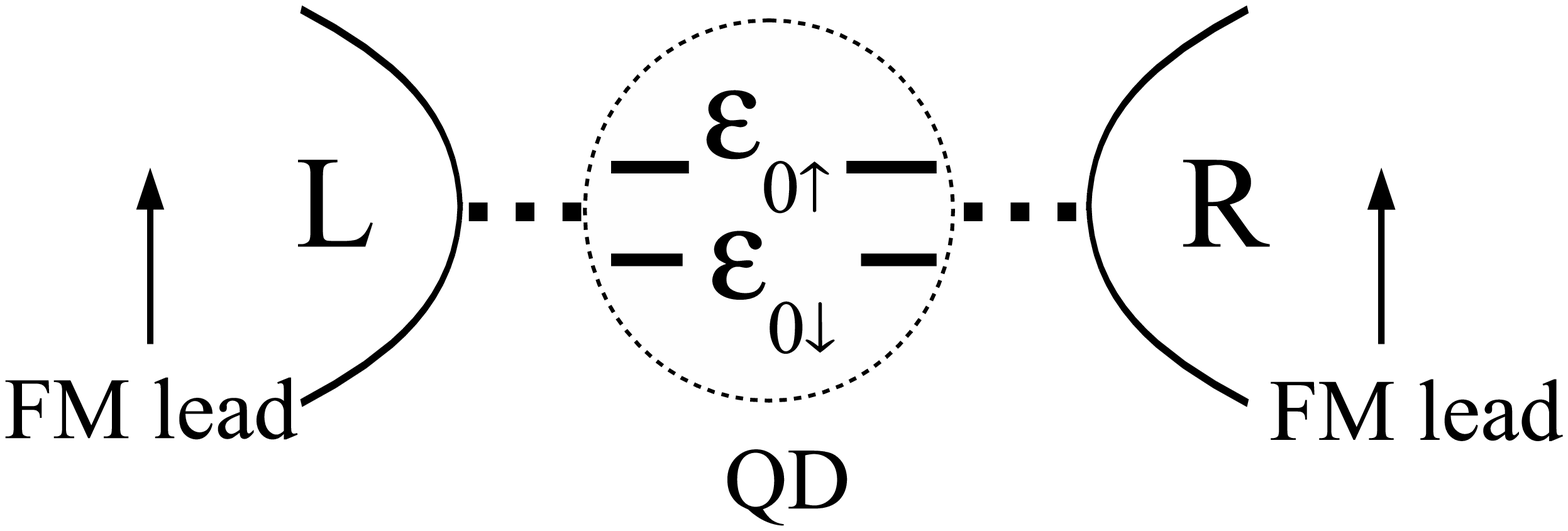}}
\caption{Schematic view of a QD connected to two full polarized ferromagnetic
leads.}
\label{fig1}
\end{figure}

The system under consideration is formed by one QD connected to two full spin-up
polarized ferromagnetic leads, as shown schematically in Fig.\ref{fig1}. The
full system is modeled by the Anderson Hamiltonian, namely $H=H_{L}+H_{D}+H_{I}$
with
\begin{eqnarray}
H_{L} &=&\sum_{i}\sum_{\sigma =\uparrow \downarrow }\varepsilon _{\sigma
}\,c_{i\sigma }^{\dagger }c_{i\sigma }-v\sum_{\langle i\neq j\rangle
}\sum_{\sigma =\uparrow \downarrow }\,\left( c_{i\sigma }^{\dagger
}c_{j\sigma }+c_{j\sigma }^{\dagger }c_{i\sigma }\right) \ ,  \nonumber \\
H_{D} &=&\sum_{\sigma =\uparrow \downarrow }\varepsilon _{0\sigma }d_{\sigma
}^{\dagger }d_{\sigma }+Un_{d\uparrow }n_{d\downarrow }\ ,  \nonumber \\
H_{I} &=&-V_{0}\sum_{\sigma =\uparrow \downarrow }\left( d_{\sigma
}^{\dagger }c_{1\sigma }+c_{1\sigma }^{\dagger }d_{\sigma }\right)
-V_{0}\sum_{\sigma =\uparrow \downarrow }\left( d_{\sigma }^{\dagger
}c_{-1\sigma }+c_{-1\sigma }^{\dagger }d_{\sigma }\right)   \nonumber \\
&-&\sum_{\sigma ,\sigma ^{\prime }=\uparrow \downarrow }t_{so}\left[ \sigma
_{x}\right] _{\sigma \sigma ^{\prime }}\left( d_{\sigma }^{\dagger
}c_{1\sigma ^{\prime }}+c_{1\sigma }^{\dagger }d_{\sigma ^{\prime }}\right)
\nonumber \\
&-&\sum_{\sigma ,\sigma ^{\prime }=\uparrow \downarrow }t_{so}\left[ \sigma
_{x}\right] _{\sigma \sigma ^{\prime }}\left( d_{\sigma }^{\dagger
}c_{-1\sigma ^{\prime }}+c_{-1\sigma }^{\dagger }d_{\sigma ^{\prime
}}\right) ,  \label{Hamiltonian}
\end{eqnarray}%
where $c_{i\sigma }^{\dagger }$ is the creation operator of an electron at site
$i$ of the leads in the $\sigma $ spin state ($\sigma =\uparrow ,\downarrow $),
and $d_{\sigma }^{\dagger }$ is the corresponding operator of an electron with
spin $\sigma $ of the QD. Moreover $n_{d\sigma }=d_{\sigma }^{\dagger }d_{\sigma
}$ and $V_{0}$ is the coupling between the QD and the leads. $U$ is the Coulomb
coupling and it will be neglected hereafter.  The potential of the wire is taken
to be zero and the hopping in the wire is $-v$. Furthermore we set the the site
energies in the leads and energy levels of the QD as
$\varepsilon
_{\sigma }=\Delta \left[ \sigma _{z}\right] _{\sigma \sigma }$ and $%
\varepsilon _{0\sigma }=\varepsilon _{0}+\mu B\left[ \sigma _{z}\right]
_{\sigma \sigma }$, respectively, where $\vec{\sigma}=(\sigma _{x},\sigma
_{y},\sigma _{z})$ is the Pauli matrix vector, $\Delta $ is the
ferromagnetic energy, $B$ is the magnetic field in the quantum dot and $%
\varepsilon _{0}$ is the energy level in the dot without magnetic field.

The stationary states of the Hamiltonian $H$ can be written as
\begin{equation}
\left\vert \psi _{\sigma }\right\rangle =\sum_{j=-\infty ,\,j\neq 0}^{\infty
}a_{j\sigma }\left\vert j\right\rangle +b_{\sigma }\left\vert 0\right\rangle
\, ,
\end{equation}
where $a_{j\sigma }$ and $b_{\sigma }$ are the probabilities amplitudes to
find the electron at the site $j$ or at the QD respectively, with energy $%
\omega =\varepsilon _{\uparrow }-2v\cos k$ or $\omega =\varepsilon
_{\downarrow}-2v\cosh \kappa $, where $\varepsilon _{\uparrow
}=2+\Delta$ and $\varepsilon _{\uparrow }=2-\Delta$. These
amplitudes obey the following linear difference equations
\begin{eqnarray}
(\omega -\varepsilon _{\sigma })a_{j\sigma }& =&-v(a_{j+1,\sigma
}+a_{j-1,\sigma })\,,\quad j\neq -1,0,1\,,  \nonumber \\
(\omega -\varepsilon _{\sigma })a_{-1\sigma }& =&-va_{-2\sigma
}-V_{0}b_{\sigma }-t_{so}\left[ \sigma _{x}\right] _{\sigma \overline{\sigma
}}b_{\overline{\sigma }}\,,  \nonumber \\
(\omega -\varepsilon _{\sigma })a_{1\sigma }& =&-va_{2\sigma
}-V_{0}b_{\sigma }\,-t_{so}\left[ \sigma _{x}\right] _{\sigma \overline{%
\sigma }}b_{\overline{\sigma }},  \nonumber \\
(\omega -\widetilde{\varepsilon }_{0\sigma })b_{\sigma }&
=&-V_{0}(a_{1,\sigma }+a_{-1,\sigma })-\,t_{so}\left[ \sigma _{x}\right]
_{\sigma \overline{\sigma }}(a_{1,\overline{\sigma }}+a_{-1,\overline{\sigma}})\ .
\label{e:diferencias}
\end{eqnarray}
where $\widetilde{\varepsilon }_{0\sigma }$ is the renormalized energy level
of the QD with spin $\sigma $,
\begin{eqnarray}
\widetilde{\varepsilon }_{0\uparrow } &=&\varepsilon _{0}+ \mu B  \nonumber \\
\widetilde{\varepsilon }_{0\downarrow} &=&\varepsilon _{0}- \mu B
\end{eqnarray}

In order to study the solutions of the above equations, we assume that
spin-up electrons are described by a plane wave with unitary incident
amplitude, $r$ and $t$ being the reflection and transmission amplitudes.
Thus we get
\begin{eqnarray}
a_{j\uparrow }& =&e^{ik\,j}+re^{-ik\,j}\,,\,j<0\,,  \nonumber \\
a_{j\uparrow }& =t&e^{ik\,j}\,,\qquad j>0\,.  \nonumber \\
a_{j\downarrow}& =&Ae^{\kappa \,j}\,,\,j<0\,  \nonumber \\
a_{j\downarrow}& =&Be^{-\kappa \,j}\,,\,j>0\,  \label{e:solut}
\end{eqnarray}
Inserting this solution in the equation of motion, we get an inhomogeneous
system of linear equations for the unknowns $t$, $r$, $A$ and $B$, leading
to the following expression for the transmission amplitude $t,$
\begin{eqnarray}
t &=&2i\alpha_{-} \sin k\left[(\omega - \widetilde{\varepsilon }%
_{0\downarrow})V_{0}^{2} -(\omega-\widetilde{\varepsilon}_{0\uparrow
})t_{so}^{2}- 2\alpha_{-} ^{2}ve^{-\kappa} \right]  \nonumber \\
&&/\Big[V_{0}^{2}\left( \omega -\widetilde{\varepsilon }_{0\uparrow}
+2\alpha_{-} e^{ik}\right) \left(\omega-\widetilde{\varepsilon }%
_{0\downarrow}-2\alpha_{-}e^{-\kappa}\right)  \nonumber \\
&+&t_{so}^{2}\left( \omega -\widetilde{\varepsilon }_{0\downarrow
}+2\alpha_{-} e^{ik}\right)\left( -\omega +\widetilde{\varepsilon}%
_{0\uparrow}+2\alpha_{-}e^{-\kappa}\right) \Big]
\end{eqnarray}
with $\alpha _{\pm }\equiv \left( V_{0}^{2}\pm t_{so}^{2}\right) /v$\,.

Notice that the above expression reduces to a single resonance when the
Rashba spin-orbit coupling is neglected ($t_{so}=0$), namely
\begin{equation}
t=\frac{i\Gamma _{0}}{\left( \omega -\widetilde{\varepsilon }_{0\uparrow
}-\Lambda _{0}\right) +i\Gamma _{0}}\ ,
\end{equation}%
where $\Gamma _{0}=2$ $V_{0}^{2}\sin k,$is the width of the resonance
centered at $\widetilde{\varepsilon }_{0\uparrow }+\Lambda _{0}$, and $%
\Lambda _{0}=\left( 2V_{0}^{2}/v^{2}\right) \Delta ,$ is the shift due to
the coupling of the QD with the leads. For $V_{0}=0$ the transmission
amplitude is also reduced to a single resonance with width $\Gamma _{so}=2$ $%
t_{so}^{2}\sin k$ and centered at $\widetilde{\varepsilon }_{0\downarrow
}+\Lambda _{so}$, where $\Lambda _{so}=\left( 2t_{so}^{2}/v^{2}\right)
\Delta ,$ is the shift due to the spin-orbit coupling of the QD with the
leads,
\begin{equation}
t=\frac{i\Gamma _{so}}{\left( \omega -\widetilde{\varepsilon }_{0\downarrow
}-\Lambda _{so}\right) -i\Gamma _{so}}\ .
\end{equation}%
For this expression we can conclude that the Rashba spin-orbit coupling
opens a new channel for tunneling through the QD. Finally, the conductance
for spin-up electrons is calculated by means of the Landauer formalism at
zero temperature
\begin{equation}
\mathcal{G}_{\uparrow }=\frac{e^{2}}{h}\,T\left( E_{F}\right) \ .
\end{equation}

\section{Result}

Evaluating the transmission probability at $\omega =E_{F}=0$ we obtain the
spin-dependent conductance
\begin{equation}
\mathcal{G}_{\uparrow}=\frac{e^{2}}{h}\,\frac{4\sin ^{2}k_{F}\left[ \alpha
_{-}\left( \varepsilon _{0}+\xi _{-}\right) -\mu B\alpha _{+}\right] ^{2}} {%
\left\vert \left( \varepsilon _{0}-\xi _{-}\right) \left( \varepsilon
_{0}-\xi _{+}\right) +2\alpha _{+}\mu B\left( e^{ik_{F}}+e^{-\kappa
_{F}}\right) -\mu^2 B^{2}\right\vert ^{2}}
\end{equation}%
where $\xi _{+}=2\alpha _{-}\cos k_{F},\xi _{-}=2e^{-\kappa _{F}}\alpha _{-}$%
, $k_{F}=\cos ^{-1}(1-\Delta /2v)$ and $\kappa _{F}=\cosh ^{-1}(1+\Delta /2v)
$. In what follows we present results for the conductance for $V_{0}/2v=0.14$
and $\Delta /2v=0.1$.

\begin{figure}[h]
\centerline{\includegraphics[angle=0,scale=0.5,clip]{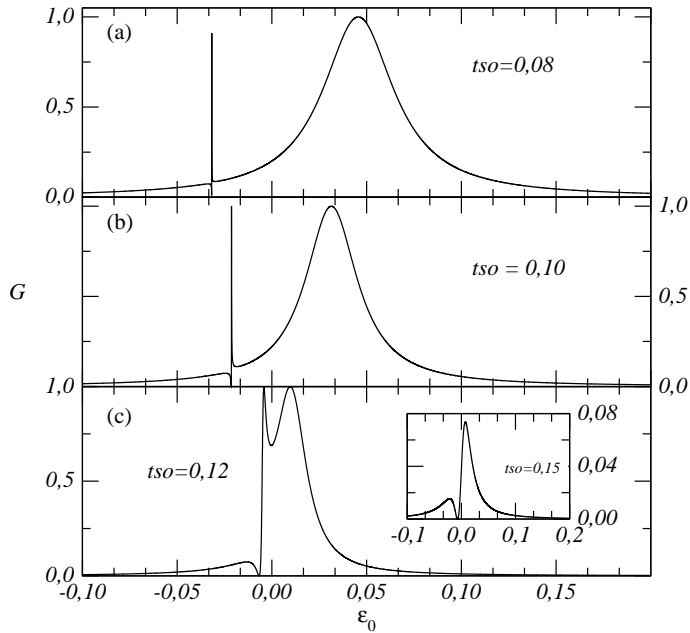}}
\caption{Spin-dependent conductance as a function of the gate voltage $%
\protect\varepsilon _{0}$ for different values of the spin-orbit coupling,
when $\protect\mu B/2v=0.001$.}
\label{fig2}
\end{figure}

Figure~\ref{fig2} displays the spin-dependent linear conductance versus the
gate voltage $\varepsilon _{0}$ for different values of the spin-orbit
coupling at fixed magnetic field. As expected, the linear conductance shows
two resonances and Fano antiresonance as a function of the Fermi energy. The
antiresonance in the conductance occurs at $\varepsilon _{0}=\xi _{-}+\mu
B\alpha _{+}/\alpha _{-}$. For small values of magnetic field ($\mu B/v\ll 1$%
) the conductance of the system can be written approximately as a
convolution of a Fano line shape and a Breit-Wigner line shape. This is,
\begin{equation}
\mathcal{G}_{\uparrow }\approx \frac{e^{2}}{h}\,\frac{\left( \epsilon
_{-}+q\right) ^{2}}{\epsilon _{-}^{2}+1}\, \frac{1}{\epsilon _{+}^{2}+1}\ .
\end{equation}
where $\epsilon_{\pm}$ are the detuning parameters measuring the energy $%
\varepsilon _{0}$ from the resonance centers and normalized by the resonance
half-width [$\epsilon _{-}=\left( \varepsilon _{0}-\xi _{-}\right) /\mu B$, $%
\epsilon _{+}=\left(\varepsilon _{0}-\xi _{+}\right) /\left( 2\alpha
_{-}\sin k_{F}\right)$] and $q=\alpha _{+}/\alpha _{-}$ is the Fano
parameter characterizing the line shape asymmetry.

\begin{figure}[h]
\centerline{\includegraphics[angle=0,scale=0.4,clip]{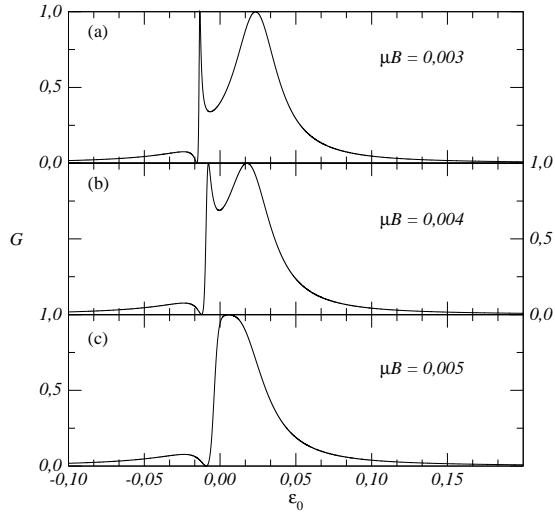}}
\caption{Spin-dependent conductance as a function of the gate voltage for
various values of the magnetic field in the dot, when $t_{so}/2v=0.1$.}
\label{fig3}
\end{figure}

The spin-dependent conductance versus the gate voltage is displayed in Fig.~%
\ref{fig3} for various values of the magnetic field in the QD. We observe
that the magnetic field modulates the spin dependent conductance, allowing
for a fine tuning of the system response. Remarkably, as the magnetic field
increases the two resonances merge into a single, broad resonance.

\begin{figure}[h]
\centerline{\includegraphics[angle=0,scale=0.4,clip]{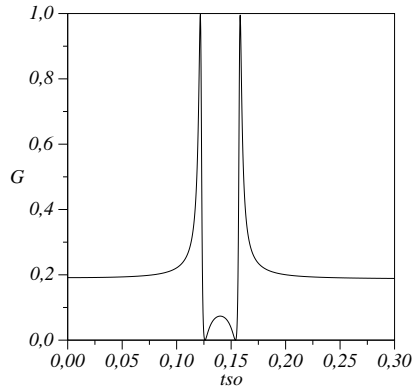}}
\caption{Spin-dependent conductance as a function of the spin-orbit coupling
$t_{so}$, when $\protect\mu B/2v=0.001$.}
\label{fig4}
\end{figure}

Figure~\ref{fig4} displays the behavior of the spin-dependent conductance as
a function of the spin-orbit coupling parameter $t_{so}$ at $\varepsilon
_{0}=0$ for different values of the magnetic field. We can see that the
conductance can also be controled by selecting a suitable value of $t_{so}$.
Notice that as $t_{so}$ is varied the conductance can pass from perfect
transmission to perfect reflection. However, $t_{so}$ is dependent on the
semiconducting materials used to fabricate the device. Therefore, in
constrat to the control gained by changing the gate voltage mentioned in the
previous paragraph, one has less control on the spin filtering capabilities
of the device by choosing different semiconductors.

\begin{figure}[h]
\centerline{\includegraphics[angle=0,scale=0.5,clip]{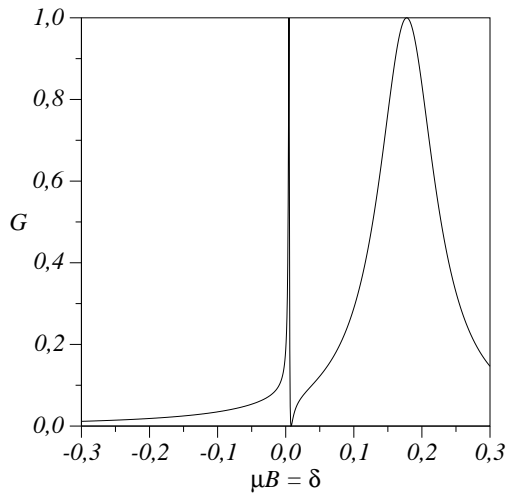}}
\caption{Spin-dependent conductance as a function of the magnetic field in
the QD, when $t_{so}/2v=0.1$.}
\label{fig5}
\end{figure}

Spin-dependent conductance at $\varepsilon _{0}=0$ as a function of the
magnetic field in the QD is displayed in Fig. \ref{fig5}. The conductance as
a function of the magnetic field also shows two resonances and one
antiresonance. The value of the magnetic field at the antiresonance is $\mu
B=2e^{-\kappa _{F}}\left( V_{0}^{2}-t_{s0}^{2}\right) ^{2}/\left(
V_{0}^{2}+t_{s0}^{2}\right)$.

\begin{figure}[t!]
\centerline{\includegraphics[width=7cm,angle=0,scale=1.0]{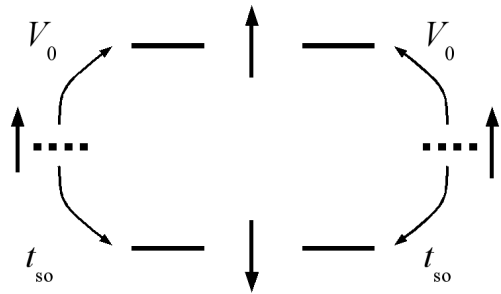}}
\caption{Schematic view of the electron tunneling through the QD.}
\label{fig7}
\end{figure}

The Rashba spin-orbit coupling opens a new channel to the conduction that
interferes with the direct channel, the key ingredient to produce the
destructive interference of the Fano effect. Figure \ref{fig7} displays a
schematic view of the electron tunneling through the QD. The electron with
spin up can tunnel directly through the level $\widetilde{\varepsilon }
_{0\uparrow }$ without spin-flip processes or also can tunnel indirectly
through the level $\widetilde{\varepsilon }_{0\uparrow }$ with two spin-flip
processes. The interference between the two tunneling paths gives rise the
Fano-Rashba effect presented in this paper.

Finally, we briefly discuss a more realistic situation when the
ferromagnetic leads are not fully polarized, as we assumed in this
work. In the case of partially polarized leads, a small fraction
of electrons with \emph{undesired} spin polarization will be
injected in the QD. Since the equations of
motion~(\ref{e:diferencias}) are linear and they lack terms
responsible of spin mixing, one could expect a gradual destruction
of the Fano-Rashba antiresonance that it would be proportional to
the loss of the polarization.

\section{Summary}

Here we investigated the electronic transport through a Rashba $QD$ with
ferromagnetic contacts. We have shown that the interference of localized
electron states with resonant electron states leads to the appearance of the
so-called Fano-Rashba effect. This effect arises from the interference of bound
levels of spin-polarized electrons in the QD with the resonant states of
opposite spin polarization. We found that the Fano-Rashba effect holds even in
presence of the electron-electron interaction in the QD.

P.\ A.\ O.\ would like to thank financial support by CONICYT/Programa
Bicen\-tenario de Ciencia y Tecnologia (CENAVA, grant ACT27). Work at Madrid was
supported by MEC (Project MOSAICO) and BSCH-UCM (Project PR34/07-15916).

\end{document}